# The role of colour preattentive processing in human–computer interaction task efficiency: a preliminary study*

R. MICHALSKI† and J. GROBELNY†

†Institute of Industrial Engineering and Management (I23),
Faculty of Computer Science and Management (W8),
Wrocław University of Technology,
27 Wybrzeże Wyspiańskiego,
50-370 Wrocław, POLAND
e-mail: Rafal.Michalski@pwr.wroc.pl
e-mail: Jerzy.Grobelny@pwr.wroc.pl

Corresponding author:

Dr. Rafał Michalski
Institute of Industrial Engineering and Management (I23),
Faculty of Computer Science and Management (W8),
Wrocław University of Technology,
27 Wybrzeże Wyspiańskiego,
50-370 Wrocław, POLAND
e-mail: Rafal.Michalski@pwr.wroc.pl
phone: +48 71 320 50 50
fax:     +48 71 320 34 32

---

* Presented in part on the XX[nd] International Seminar of Ergonomics Teachers, Miedzyzdroje, Poland 2006.

## Abstract

In this paper, results of experimental research on the preattentive mechanism in the human–computer interaction (HCI) were presented. Fifty four subjects were asked to find interface elements from various panel structures. The arrangements were differentiated by their orientation (vertical, horizontal), colour pattern (ordered, unordered) and object background colours (green–blue, green–red, blue–red). The main finding of the study generally confirms the profits provided by the visual preattentive processing of the colour feature in graphical panel operation efficiency. However, the vertical way of arranging the items in search layouts resulted in decreasing the preattentive effect related to the item background colour. In regular, chessboard-like patterns of different coloured items, the effect of the early vision was less salient than in the case of structures with randomly dispersed colours. The reported results can help in designing efficient graphical user–computer interfaces in many interactive information systems.

## Keywords

visual preattentive processing; human–computer interface design; graphical objects; dialogue windows; search task efficiency

## 1. Introduction

The term of preattentive visual information processing was introduced in the early eighties of the twentieth century by Treisman (1982). Many studies showed that there is a limited set of features that seem to be processed in parallel and appear to be able to guide the deployment of attention. The preattentive mechanism constructs some kind of meta-objects on the basis of basic features. These objects work as a guide for the human attention during the serial phase of visual searching tasks. The factors considered to invoke the preattentive procedure include, for instance colour, motion, orientation, pattern or shape of the perceived objects. On the other hand, there are of course some features that are not associated with the preattentive process. For example, Wolfe and DiMase (2003) showed that the lines intersections probably cannot be regarded as such a feature. A broad range of attributes that might guide the deployment of attention were presented and grouped by the likelihood of their occurrence by Wolfe and Horowitz (2004). The latest research of Bichot and colleagues (2001, 2005) showed that preattentive mechanisms can be observed on the neurophysiological level – in the cerebral cortex. In their experiments, a macaque monkey was learned to choose coloured objects from a set of various distractors. The activation of different layers of visual cortex neural cells was recorded. Two separate phases were recognised in the activation plots. The first connected with colour identification during preattentive processing and the second attributed to the moment of finding the searched figure. The role of the preattentive processing is undoubtedly important in many day-to-day activities and, of course, is of great significance in the field of Human–Computer Interaction (HCI). Some of the screen design recommendations proposed in the form of well known Gestalt Theory laws (Change et al., 2002) such as proximity, similarity or closure are related to display attributes that have the early vision nature. In the presented study we focus on the human–computer dialogue style named direct manipulation and its 'point and click' method (Shneiderman, 1982, 1983). Naturally, this method involves very often visual search for the specific item. Pointing various graphical elements on the screen and confirming (usually by clicking) the execution of a given activity requires the use one of available peripheral devices, such as light pens, joysticks, touch screens or computer mice (Greenstein and Arnaut, 1988). Until now the direct manipulation performed by means of the computer mouse is probably one of the most popular ways of interacting with computers. This interaction style is used by millions of users in many operating system all over the world.

## 2. Related research

The research concerned with the human performance in a direct manipulation interaction style where arrangements of graphical objects are presented, may be divided into three main trends (Michalski et al., 2006). The first one includes investigations of the movement time in the 'point and click' method. In this case, the user selects a given, constantly visible, graphical object by means of a pointing device. These types of studies do not take into account the process of searching for an item and the measured time reflects only the visually controlled motor activity. The research in this area started in the 1960's (e.g., English et al., 1967). In the 1970's, the classical Fitts law was used to describe the process of pointing graphical objects on the computer screen (Fitts, 1954; Fitts and Peterson, 1964). A comprehensive review of Fitts law application for various pointing devices, presented MacKenzie (1991, 1992) and later Plamondon and Alimi (1997). The works in this trend are still conducted in standard and new interaction styles, for instance Murata and Iwase (2001) or Soukoreff and MacKenzie (2004).

In subsequent studies, it occurred that the Fitts law is insufficient for explaining the results related to the process of choosing items on a computer screen and cognitive components should also be taken into consideration (Card et al., 1983). In the second trend, the subjects had to find a particular graphical object among the group of distractors in changing circumstances. These investigations did not require the participant to point the specified element. Many works were published in this field, however one of the first focused on graphical interface features was conducted by Backs and colleagues (1987). They asked the users to find a target object in various vertical and horizontal menus and report an associated numerical value. Other investigations regarded the visual search within randomly generated alphabetic arrays (Scott and Findlay, 1991), the effect of icon presence and their grouping on the process of searching a file (Niemelä and Saarinen, 2000) or the spatial menu layout in web pages and its impact on the visual search performance (Schaik and Ling, 2001). Also some studies were carried out by means of eyeball tracking systems. Näsänen and colleagues (2001) analyzed visual search times of Latin characters arranged in a square matrix, depending on the element's contrast and matrix size. In turn, Näsänen and Ojanpää (2003) investigated the effects of graphical object contrast and sharpness on the speed of searching for a specified icon.

The third area of research is a combination of the first two groups. This time the experimental task involves graphical object identification and selection by pointing and clicking on the desired item. The studies related to movement time and visual search proved that the relationship between these two factors is not always of additive nature (Hoffmann and Lim, 1997). In the HCI field this type of research is not very popular. Probably the first studies concerned with various design configurations and efficiency measures including simultaneously visual search and movement time were presented by Deininger (1960) in the 1960's. He investigated the performance of keying telephone numbers using various layouts of numerical dialing keys. Succeeding studies on keying in physical and virtual keypads or keyboards were reviewed in detail by Kroemer (2001). The research on choosing virtual graphical objects from a group was most likely initiated by Drury and Hoffmann (1992). Many aspects of the virtual keyboards were later also analysed by various researchers (e.g., Sears et al., 1993; MacKenzie and Zhang, 1999; Lee and Zhai, 2004; Li et al., 2006). Apart from the keypads and keyboards studies there was also some research focused directly on graphical computer interface features such as the menu orientation (Shih and Goonetilleke, 1998), functional icon grouping (Goldberg and Kotval, 1999), the way of symbols coding (Ramakrishnan et al., 1999) or the icon border type and quality (Fleetwood and Byrne, 2002). The article published by Michalski and colleagues (2006) is one of the latest works in this area. The authors examined the influence of a number of geometrical parameters of the 42 graphical structures on the efficiency of the 'point and click' method.

In all of the trends, the user performance can be assessed by registering the time required to complete the task and a number of incorrect selections.

## 3. Research objectives

The present investigation can be placed in the trend of research related to both movement time and visual search tasks in human–computer communication. In previous studies it was shown that the Fitts formula is inadequate in this type of search tasks and other factors such as geometrical features of the panels may substantially influence the user operation efficiency (Grobelny et al., 2005; Michalski et al., 2006). Although it is well known that the searching process may be improved by preattentive activities, we had great difficulties with finding HCI publications explicitly dealing with this perceptual mechanism. Therefore, the main purpose of this study was to find consequences of the colour and colour pattern as preattentive factors in the searching and clicking on the object in the computer graphical interface. Additionally, we wanted to check whether panel orientation effect influences the efficiency of searching and clicking a colourful object.

We used standard graphical buttons that are identical to those included in various computer programs and arranged the analysed graphical objects into groups typically employed in many applications. In previous visual search studies the stimuli were presented differently. On one hand charts or displays occupying significant part of the visual field were used (e.g., Scott and Findlay, 1991; Murata and Furukawa, 2005; Park et al., 2006) and on the other hand the search was performed within the visual lobe (e.g., Pashler, 1987; Klein and Farrel, 1989). On the basis of such investigations, it is rather hard to draw direct conclusions or predict how the people will behave in situations characteristic of human–computer interaction. In this paper, the searching area is limited to dimensions and arrangements of typical toolbars and in this sense the experimental tasks are considerably closer to real HCI conditions.

We decided to use letters and numbers instead of meaningful icons for two basic reasons. Firstly, the letters and numbers are widely known and equally easy recognizable by almost any possible computer user. If we had used a set of icons from a real software, the obtained results could have been influenced by the fact that some subjects would be more familiar with them than other users. Secondly, the chosen graphical objects can be found in a real computer programs as virtual keyboards. They are becoming more and more popular not only as a support for disabled computer users but also as one of the main input tools in Internet kiosks or small devices such as palmtops, mobile phones, etc.

We studied the mouse pointing process along with preattentive visual search which usually did not take place in previous studies. The earlier studies in this field quite often did not use pointing devices at all, even if they were carried out by means of the computer display. The investigators used for instance a keyboard to confirm the accomplishment of a task (Niemelä and Saarinen, 2000; Sears et al., 2001; Schaik and Ling, 2001; Liu et al. 2002; Pearson and Schaik, 2003; Al-Harkan and Ramadan, 2005) or an external device with a reaction button (Backs et al., 1987; Murata and Furukawa, 2005). Sometimes the researchers used the computer mouse to click on one specific confirmation button (e.g., OK) instead of pointing the searched graphical item (e.g., Näsänen et al. 2001a, 2001b; Lindberg and Näsänen, 2003;

Näsänen and Ojanpää, 2003). There were some studies in the HCI field, in which search and click methods were employed (e.g., Shih and Goonetilleke, 1998; Goldberg and Kotval, 1999; Fleetwood and Byrne, 2002; Hornof, 2004), but as far as we are aware none of the papers was concerned with typical toolbar arrangements and colour preattentive processing. From the diversity of the above procedures there comes up a question whether the inclusion of the pointing devices such as computer mice, light pens, styluses considerably influences the typical toolbar search results. There are some premises that this could take place. Namely, it has been observed that the same areas of the human brain are activated during manual response and preparing the saccade movement (Wurtz et al., 1982; Kustov and Robinson, 1996; Colby and Goldberg, 1999). These studies' results suggest that there maybe some interaction between hand movement and visual search and it can influence the HCI process as a whole. Similar conclusions may be drawn from the series of experiments on concurrent decision and movement tasks carried out by Hoffmann and Lim (1997). The authors showed that the manual–decision tasks are complex and therefore it is justified to investigate both conditions at the same time. Thus, it seems also to be useful to analyse simultaneously finding and clicking objects in graphical toolbars with the control of the colour visual preattentive processing effect.

## 4. Method

### 4.1. Participants

Overall, 54 students of Computer Science and Management Faculty from the Wrocław University of Technology participated in the experiments. All reported having a normal or corrected to normal visual acuity. The subjects were at the age of 19–25 (Mean = 23.4, Standard Deviation = 0.998) and all attended the full-time Master's program. There were fewer male participants (17 subjects, 31%) than females (37 subjects, 69%). Almost all (93%) of the volunteers worked with computers on a daily basis, the rest at least several times a week. All of the users spent more than two hours a day at a computer and there were even 6 people (11%) who spent more than 10 hours a day. The number of hours per day working with computers included only the time spent on actively using the software. Activities such as watching TV on the computer screen or listening to music being played were not taken into consideration.

### 4.2. Apparatus

A special software similar to the one utilised in the work of Michalski et al. (2006) was prepared and used to conduct the experiments. The program was developed in a MS Visual Basic™ 6.0 environment and was based on a relational MS Access™ database. The database was used to store experimental information required for further analysis such as acquisition times or errors made by subjects. The gathered data could be easily transferred to computer applications supporting statistical analysis (e.g., SPSS™, Statistica™) by means of Open DataBase Connectivity technology. The research was carried out in teaching laboratories on uniform personal computers equipped with identical optical computer mice and 17" monitors of the CRT type. The computer screen resolution was set at 1024 by 768 pixels, and default computer mouse parameters were set.

### 4.3. Independent variables

Three independent variables differentiating the analyzed graphical object structures were manipulated: the object background colour, panel colour pattern and orientation.

Graphical object background colour (OBC) was chosen because the colour it is one of the most important and undisputed component of early vision processing (Wolfe and Horowitz, 2004). Three different basic colours were employed: red, green, and blue. The red and blue were chosen as extreme wave lengths from the visible spectrum and the green as an intermediate value. Moreover, they are well known primary colours and the various combination of the three components is used to reproduce other colours in the, so called, RGB additive colour model (Wright, 1928; Guild, 1931). These colours are quite often used in various studies (e.g., Bichot et al., 2001; Liu et al., 2002; Ojanpää and Näsänen, 2003; Pearson and Schaik, 2003; Li et al., 2007). In this study they were used in pairs: red–blue (RB), red–green (RG), and green–blue (GB).

Two different panel colour patterns (PCP) were employed. The first one was ordered (OR) and is similar to a chessboard and the second one unordered (UO) where two colours are mixed at random. By including the panel colour pattern factor into our research, we wanted to check whether there are any discrepancies in the acquisition times when the pattern differs. In other words, we wanted to know if the colour preattentive processing depends on the way of the coloured object arrangement within the given panel. The answer for this question seems to be especially important for the practitioners and GUI designers.

Two panel orientations (POR) were designed as typically used in the graphical interfaces: horizontal (H) and vertical (V). The panel orientation variable was used mainly from practical reasons as in the contemporary computer programs the vertical and horizontal configurations of graphical objects are very common. Though this variable was already studied, the results were ambiguous (Deininger, 1960; Backs et al., 1987; Scott and Findlay, 1991; Shih and Goonetilleke, 1998; Schaik and Ling, 2001; Pearson and Schaik, 2003; Grobelny et al., 2005). Additionally, the selection of this factor allowed for comparing the obtained results directly with similar but grey panels examined by Michalski et. al. (2006).

### 4.4. Dependent measures

Two dependent variables were recorded, the operation time and the number of errors made. The time was computed from when the START button was pressed, to when the given graphical object was selected. The error occurred when the participant pointed different than the required item. The user did not receive immediate feedback about the mistake. The information about the average acquisition times and number of faults was only shown after every ten tasks were completed.

### 4.5. Experimental design

The mix of the three independent factors produced 12 different cases of searched panels: (three background object colours) × (two panel colour patterns) × (two panel orientations). A standard mixed model design (between and within subjects) was used to investigate all of the 12 sets of objects. The panel orientation was treated as a between factor, whereas the other two variables: panel colour patterns, and background colours of items were tested within. The volunteers were randomly divided into two groups. One group consisting of 28 persons examined only horizontal panels, the second (26 subjects) – the vertical ones. All of the analysed panels consisted of 36 identical buttons (Figure 1). Graphical objects representing 26 Latin alphabet characters and ten Arabic numbers were placed on these buttons. The bolded Times New Roman font type in a size of 12pt was employed. The numbers from one to ten were utilised in order to avoid potential mistakes between the O character and a 0 (zero) digit. The standard square buttons used in Microsoft® operating systems were employed. The side button size in TWIPs (At 1024 by 768 screen resolution, one pixel amounts to 15 TWIPs), pixels, millimetres and visual angle amounted to 330, 22, 6, 0°41' respectively. These types of elements are commonly utilised in many computer programs, e.g., in a popular MS Office™ package. The choice of a square shape of elements was based on the findings that the square items are better operated on the computer screen than rectangular shapes (Martin, 1988). The subjects were asked to select a graphical item from a panel containing randomly placed buttons. The investigated arrangements were located in the upper left screen corner and were moved away from the screen edges by 270 TWIPs in order to minimise the effect of easier selection of items placed on the external borders of the sets. The distance between the screen border and the panels was equalled to the height of the top title bar used in most Microsoft® operating systems dialogue windows. The whole graphical structure was visible exclusively during the visual search process. The effects of learning were not examined. The direct manipulation tasks were executed by means of a standard computer mouse. The distance between the user and the computer monitor was set approximately at 50 cm. The visual angles of the examined horizontal layouts were 12.9° × 1.4° and for vertical ones – 1.4° × 12.9°.

A horizontal, ordered, green–blue panel (H_OR_GB)

A horizontal, unordered, red–blue panel (H_UO_RB)

A vertical, unordered, red–green panel (V_UO_RG)

Figure 1. Exemplary panel configurations examined in the experiment.

## 4.6. Procedure

The employed procedure was similar to the one used in the work of Michalski et al. (2006). The subjects were informed about a purpose and a detailed range of the study. The study began by filling out a questionnaire regarding personal data and computer literacy. Just before the experiment, each subject executed test tasks that were not recorded. Attemptive trials were performed until the user stated that he was ready to start the real tasks. Then a dialogue window appeared with a graphical object, a START button, and an order of searching a specified object – and at this moment the tested graphical panel was not visible. When a user clicked the START button, the instruction window disappeared and one of the examined structures was shown. Subjects were to find and select the earlier presented graphical object as quick as possible. The START button appeared for each trial so every time, the participant had to click START first and then the searched item. During the experiments, only research software was available and visible, and clicking anything else than the START button, produced a message box with guiding information. Volunteers were allowed to use only a computer mouse during the efficiency test. All other input devices were programically disabled. For every single panel arrangement, there were ten execution orders with a randomly chosen button. Once shown, an element could appear again. The order of presenting an individual group of items was set at random for every user. The location of every graphical object within the given panel was also randomly specified. Information about obtained mean acquisition times and a number of incorrect attempts was shown every 10 trials.

## 5. Results

### 5.1. Descriptive statistics

The basic descriptive parameters of the dependent variable were grouped in three main categories: central tendency measures, variability measures, and shape characteristics and are presented in Tables 1, 2, and 3. The data in Table 1, shows that the median, and mean values differ considerably one from another. The mean is as much as 22% bigger than the median value. In the case of the Gaussian distribution, these three parameters should be comparable. Also the calculated skewness (4.31) and the kurtosis values (42.5) were noticeably different from these parameters values characteristic of the normal distribution. The positive sign of the skewness denotes that most of the variate values are located on the left hand side of the distribution and the large kurtosis value estimated from the sample suggests that the probability density distribution in this case is markedly less dispersed than the normal distribution.

Table 1. Central tendency measures of acquisition times.

| Central tendency measures | |
|---|---|
| Mean | 2 118 ms |
| Geometric mean | 1 820 ms |
| Harmonic mean | 1 615 ms |
| Median | 1 743 ms |

Table 2. Variability measures of acquisition times.

| Variability measures | |
|---|---|
| Minimum | 609 ms |
| Maximum | 28 579 ms |
| Standard deviation | 1 478 ms |
| Variance | 2 185 245 $(ms)^2$ |
| Variability coefficient | 70 % |

Table 3. Shape characteristics of acquisition times.

| Shape characteristics | |
|---|---|
| Skewness | 4.31 |
| Kurtosis | 42.5 |

The main statistical characteristics of 'search and point' times obtained for all experimental conditions are presented in Table 4. From these results, it may be noticed that characteristics for individual examined panels are very similar in their structure to the parameter computed for the whole acquisition time variate.

The presented data showed that the horizontal unordered configuration consisting of items in red and blue (1 908 ms) was operated the fastest. The worst mean results were obtained for the ordered vertical arrangements containing red and blue objects (2 343 ms). The percentage discrepancy between the best and the worst result was equal to 23%. The statistical significance of differences between the independent variables is analysed in the next section.

Table 4. Acquisition times characteristics for individual panels.

| No. | Panel type | N | Median (ms) | Mean (ms) | Standard error (ms) | Standard deviation (ms) | Minimum (ms) | Maximum (ms) |
|---|---|---|---|---|---|---|---|---|
| 1. | H_OR_GB | 276 | 1 679 | 2 238 | 103 | 1 718 | 688 | 11 750 |
| 2. | H_OR_RB | 275 | 1 579 | 2 059 | 126 | 2 097 | 625 | 28 579 |
| 3. | H_OR_RG | 276 | 1 649 | 1 940 | 68 | 1 130 | 625 | 9 422 |
| 4. | H_UO_GB | 275 | 1 625 | 2 077 | 97 | 1 602 | 703 | 12 437 |
| 5. | H_UO_RB | 278 | 1 594 | 1 908 | 75 | 1 243 | 609 | 13 360 |
| 6. | H_UO_RG | 274 | 1 625 | 1 979 | 76 | 1 258 | 640 | 10 125 |
| 7. | V_OR_GB | 257 | 1 875 | 2 289 | 89 | 1 428 | 703 | 10 313 |
| 8. | V_OR_RB | 258 | 1 938 | 2 343 | 89 | 1 430 | 719 | 10 204 |
| 9. | V_OR_RG | 258 | 1 875 | 2 331 | 103 | 1 657 | 625 | 15 188 |
| 10. | V_UO_GB | 255 | 1 734 | 2 052 | 71 | 1 135 | 656 | 7 610 |
| 11. | V_UO_RB | 259 | 1 766 | 2 055 | 77 | 1 233 | 718 | 7 672 |
| 12. | V_UO_RG | 259 | 1 875 | 2 177 | 88 | 1 420 | 703 | 11 157 |

## 5.2. Analysis of variance

Taking into account the descriptive characteristics of the acquisition time distribution presented in the previous section, especially big values of the skewness and kurtosis, all of the analyses of variance were conducted by means of the Generalised Linear Models (GZLM). It was also assumed that the dependent variable has the inverse Gaussian (IG) distribution. The GZLM was first defined by Nelder and Wedderburn (1972), and as opposed to General Linear Models (GLM), do not require a dependent variable to have a normal distribution. Additionally, the assumption about constant variance of a random component need not to be met. The GZLM incorporate an analysis of variance and require statistical distribution of a dependent variable to belong to a natural exponential family of distributions. In the work of Michalski (2005), it was shown that the hypothesis about the IG character of the acquisition time empirical distribution for the types of graphical structures comparable with the examined in this study cannot be rejected.

A three factorial analysis of variance based on the GZLM was employed for assessing the effects of the graphical object background colour, the panel colour pattern the and the panel orientation. The results of the ANOVA are shown in Table 5.

Table 5. GZLM analysis of variance results.

| Effect | df | Wald statistics (W) | p |
|---|---|---|---|
| Panel orientation (POR) | 1 | 17.6 | *0.000028 |
| Panel colour pattern (PCP) | 1 | 14.0 | *0.00018 |
| Object background colour (OBC) | 2 | 2.4 | 0.30 |
| POR × PCP | 1 | 2.3 | 0.13 |
| POR × OBC | 2 | 8.2 | *0.016 |
| PCP × OBC | 2 | 3.2 | 0.20 |
| POR × PCP × OBC | 2 | 0.3 | 0.85 |

* The results significant at a level 0.05

According to the obtained results, the effects of panel orientation and panel colour pattern were statistically significant whereas the object background colour had no meaningful impact on the obtained acquisition times. The basic statistics related to these factors are presented in Tables 6 and 7 and illustrated in Figure 2. Because of the different than normal character of the dependent variate minima, maxima and medians were also included. The means depending on the panel orientation were lower for horizontal arrangements than for vertical ones and the difference amounted to 175 ms (8.6%). The analysis indicated also that unordered layouts were operated better by 7.7% (157 ms) than ordered structures.

Table 6. Results for the panel orientation (POR) factor (df = 1, W = 17.6, p = 0.000028).

| No. | POR | N | Median (ms) | Mean (ms) | Standard error (ms) | Standard deviation (ms) | Minimum (ms) | Maximum (ms) |
|---|---|---|---|---|---|---|---|---|
| 1. | Horizontal | 1 654 | 1 625 | 2 033 | 38 | 1 546 | 609 | 28 579 |
| 2. | Vertical | 1 546 | 1 860 | 2 208 | 36 | 1 397 | 625 | 15 188 |

Table 7. Results for the panel colour pattern (PCP) factor (df = 1, W = 14.0, p = 0.00018).

| No. | PCP | N | Median (ms) | Mean (ms) | Standard error (ms) | Standard deviation (ms) | Minimum (ms) | Maximum (ms) |
|---|---|---|---|---|---|---|---|---|
| 1. | Ordered | 1 600 | 1 781 | 2 196 | 40 | 1 612 | 625 | 28 579 |
| 2. | Unordered | 1 600 | 1 688 | 2 039 | 33 | 1 327 | 609 | 13 360 |

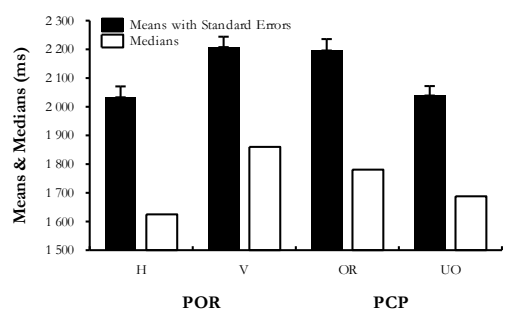

Figure 2. Results for the panel orientation (df = 1, W = 17.6, p = 0.000028) and panel colour pattern (df = 1, W = 14.0, p = 0.00018) factors.

From among the possible interactions in the ANOVA only the interaction of (panel orientation) × (object background colour) was significant. Detailed results are shown in Table 8 and Figure 3. It can be observed that within the given panel background colours, the means for the panel orientation effect were similar only in the case of green–blue level – merely 14 ms (0.7%). For the remaining two variants these discrepancies were decidedly bigger (red–blue – 11% and red–green – 15%).

Table 8. Results for the interaction between panel orientation (POR) and object background colour (OBC) factors (df = 2, W = 8.2, p = 0.016).

| No. | POR × OBC | N | Median (ms) | Mean (ms) | Standard error (ms) | Standard deviation (ms) | Minimum (ms) | Maximum (ms) |
|---|---|---|---|---|---|---|---|---|
| 1. | H_GB | 551 | 1 641 | 2 157 | 71 | 1 661 | 688 | 12 437 |
| 2. | H_RB | 553 | 1 594 | 1 983 | 73 | 1 722 | 609 | 28 579 |
| 3. | H_RG | 550 | 1 633 | 1 960 | 51 | 1 195 | 625 | 10 125 |
| 4. | V_GB | 512 | 1 821 | 2 171 | 57 | 1 294 | 656 | 10 313 |
| 5. | V_RB | 517 | 1 875 | 2 199 | 59 | 1 342 | 718 | 10 204 |
| 6. | V_RG | 517 | 1 875 | 2 254 | 68 | 1 543 | 625 | 15 188 |

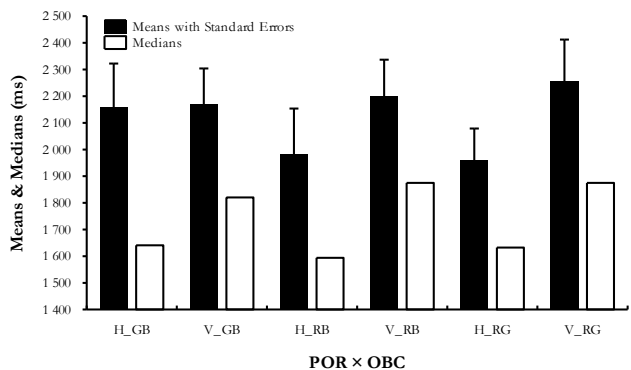

Figure 3. Results for the interaction between panel orientation (POR) and object background colour (OBC) factors (df = 2, W = 8.2, p = 0.016).

In order to provide more insights with respect to this interaction's nature, a series of post-hoc one-way GZLM analyses of variance were employed. It occurred that the object background colour factor was insignificant for vertically oriented layouts (df = 2, W = 1.3, p = 0.53) and statistically meaningful in the case of horizontal structures (df = 2, W = 9.3, p = 0.0094). Additional one-way ANOVAs on two levels showed also that the differences between the object background colour factor levels in horizontal arrangements were significant for two pairs: green–blue and red–blue (df = 1, W = 5.8, p = 0.016) as well as green–blue and red–green (df = 1, W = 8.2, p = 0.0043). The discrepancy for the pair red–blue and red–green was not statistically meaningful (df = 1, W = 0.1, p = 0.72). Further investigations revealed that in green and blue panels no considerable differences were observed between vertical and horizontal orientations (df = 1, W = 0.031, p = 0.86). The panel orientation effect was significant for red–blue (df = 1, W = 9.2, p = 0.0024) and red–green (df = 1, W = 17, p = 0.000026) types of graphical structures.

### 5.3. Errors analysis

A total of 40 errors were made by subjects during the experiments, which accounts for 1.25% of all executed orders. The biggest proportion of wrong selections (2.1%) was noted for the unordered horizontal panel with red and green objects (H_UO_RG). The lowest number of errors (0.4%) was registered during testing unordered vertical arrangements consisting of red–blue and red–green items. A nonparametric Chi-square test was used to verify the significance of differences in observed incorrect selections for analysed effects. The obtained results are put together in Table 9.

Table 9. Error results analysis.

| Effect | df | $\chi^2$ | p |
|---|---|---|---|
| Object background colour | 2 | 1.9 | 0.39 |
| Panel colour pattern | 1 | 0.0 | 1.0 |
| Panel Orientation | 1 | 2.8 | 0.094 |

The results demonstrated no statistically meaningful ($\alpha = 0.05$) differences in the number of errors made for the specified factors. The biggest discrepancy was observed in the case of the panel orientation effect ($p = 0.094$). The percentage of mistakes for vertical layouts amounted to 0.9% and for horizontal – 1.5%.

## 6. Discussion

### 6.1. ANOVA results

The main goal of this study was to investigate the role of the basic preattentive mechanisms combined with some graphical interface characteristics on human-computer interaction task efficiency. It was found that in the analysed type of user tasks the panel colour pattern and its orientation happened to be statistically significant while the graphical object background colour as a whole did not influence operation efficiency. Furthermore, some evidence of statistically meaningful interaction between the panel orientation and object background colour were provided. The considerable influence of the object background colour factor on the task efficiency was noticed among the horizontal layouts.

### 6.1.1. Panel orientation

A general predominance of the horizontal arrangements over vertical ones was observed. This fact was also reported in previous experiments by various researchers, however the results in this respect were not always consistent. For example, Backs and colleagues (1987) reported significantly faster search times in vertical than in horizontal layouts. It should be noted that in their investigation, accomplishing the task for horizontal structures was more difficult than in the case of vertical configurations. The value associated with a horizontal menu item could be mistaken with the other, adjacent menu text, whereas for vertical sets such a situation was rather impossible. In the works of Scott and Findlay (1991) and Shih and Goonetilleke (1998) horizontal configurations were operated faster than the vertical ones. In the latter case, additionally, this outcome was not dependent on the language (Chinese or English). Schaik and Ling (2001) obtained faster search times for menus located at the top or left of the screen, but very similar studies conducted by Pearson and Schaik (2003) yielded shorter acquisition times only for horizontal menu locations. In experiments analysed in the Grobelny et al. (2005) investigation, the horizontal structures were better than vertical in terms of efficiency. There could be some possible explanations for the better user performance during testing horizontal layouts. They can be, to some extent, accounted for by the culturally conditioned habit of analysing objects horizontally from the left to the right hand side. Also the popularity of horizontally shaped menus and toolbars in numerous and widely spread computer programs may have an impact on the results. However, the most influential factor perhaps lies in the shape of the human's field of view which resembles a horizontal ellipse for various wave lengths.

### 6.1.2. Panel colour pattern

From among two patterns used for placing colourful objects on the examined panels, the random disposition occurred to be more helpful in executing the search and select tasks. In the case of the chessboard pattern, the colour preattentive processing effect was decidedly less distinct and unexpectedly the difference between mean acquisition times was statistically significant.

### 6.1.3. Graphical object background colour

Although the main ANOVA showed irrelevant influence of the object background effect, the additional analysis of the statistically significant interaction of panel orientation and object background colour factors demonstrated that the object background colour effect considerably affected the operation mean times in horizontal structures. The significant differences between individual levels of this effect can be possibly attributed to the dissimilar contrasts applied in examined panels. The less salient effect of background object colour for green–blue panels in comparison with the other pairs of colours, can be probably explained by the lower colour contrast. The difference between red–blue and red–green arrangements, was statistically of no importance due to the similar reason: comparable contrast levels of these two pairs of colours.

Despite the unbalanced ANOVA, we also compared the earlier results gathered by Michalski and colleagues (2006) with the data registered during this examination. We found that mean operation times were decidedly shortest for colourful sets than for grey panels (df = 1, W = 57, p < 0.000001). The difference between the average values of operation times amounted to 317 ms (15%). Also the discrepancies in standard deviations (SD) and mean standard errors (MSE) were substantial. The values of SD and MSE for the layouts used in the work (Michalski et al., 2006) were bigger by 52% (765 ms) and 114% (30 ms) respectively, in comparison with these values computed in the present examination.

After conducting a series of additional one-way GLZM ANOVAs, it occurred (Table 10, 11 and Figures 4, 5) that horizontal coloured layouts were better than grey panels in all variants excluding the horizontal green and blue chessboard pattern configuration (H_OR_GB). The average times of searching and selecting a target were decidedly shorter in the case of colourful, unordered vertical arrangements (V_UO) than homogeneously grey panels (V_GR). On the other hand, the difference in mean times between vertical grey (V_GR) and vertical ordered colourful configurations was statistically significant only for the green–blue pair of colours (V_OR_GB).

Table 10. GLZM one-way ANOVAs results for horizontal panels (df = 1)

| Colourful panel type | Grey horizontal panel (H_GR) (Mean = 2 347 ms, Median = 1 763 ms, SD = 1 927 ms, MSE = 68 ms, N = 803) |
|---|---|
| H_OR_GB | W = 1.22, p = 0.270 |
| H_OR_RB | W = 9.65, *p = 0.00189 |
| H_OR_RG | W = 23.0, *p = 0.00000163 |
| H_UO_GB | W = 8.53, *p = 0.00349 |
| H_UO_RB | W = 26.7, *p = 0.000000243 |
| H_UO_RG | W = 17.3, *p = 0.0000311 |

\* The results significant at a level 0.05

Table 11. GLZM one-way ANOVAs results for vertical panels (df = 1)

| Colourful panel type | Grey vertical panel (V_GR) (Mean = 2 522 ms, Median = 1 902, SD = 2 518 ms, MSE = 89 ms, N = 801) |
|---|---|
| V_OR_GB | W = 4.70, *p = 0.0302 |
| V_OR_RB | W = 2.65, p = 0.103 |
| V_OR_RG | W = 2.97, p = 0.0847 |
| V_UO_GB | W = 22.9, *p = 0.00000169 |
| V_UO_RB | W = 22.5, *p = 0.00000205 |
| V_UO_RG | W = 11.1, *p = 0.000847 |

\* The results significant at a level 0.05

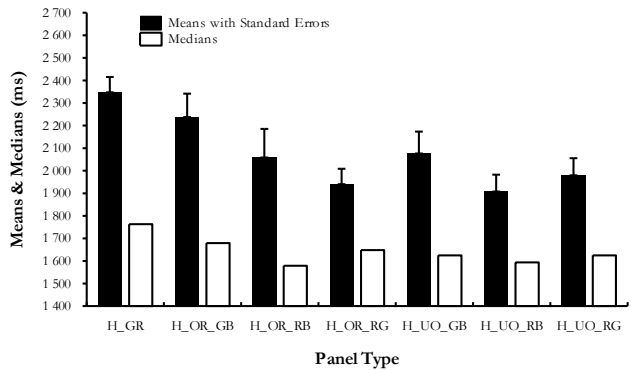

Figure 4. Acquisition times for horizontal, coloured and grey panels.

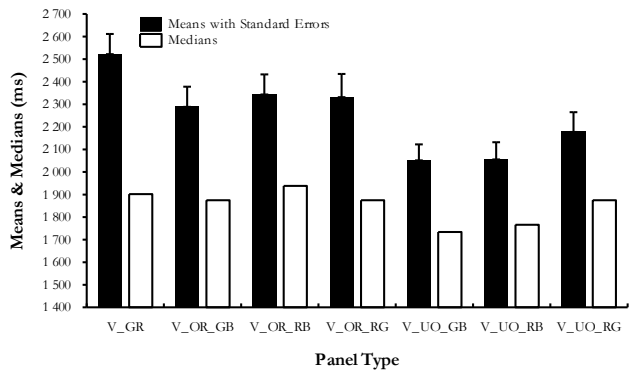

Figure 5. Acquisition times for vertical, coloured and gray panels.

These results indicate that the use of various background object colours in most cases enables a user to search faster for particular items. This could probably be attributed to the preattentive properties of the colour feature.

## 6.2. Errors

A total number of incorrect selections was not big and the examined effects did not affect the measured errors. The obtained accuracy (1.25%) is close to the faults rates reported in other studies. For instance, in the works of Grobelny and colleagues (2005), Schaik and Ling (2001) the error rate did not exceeded 2%, while in studies of Backs et al. (1987), Shih and Goonetilleke (1998), Pearson and Schaik (2003), Michalski et al. (2006) the failures occurred in less than 3% of all trials. There were also investigations, in which the incorrect response rate was decidedly higher, for example 4.8% reported by Scott and Findlay (1991) or even 7.5% obtained by Simonin et al. (2005). These discrepancies can be, in part, credited to diverse experimental conditions used in these works. In the experiments presented by Michalski and others (2006) the error rate observed for grey panels having the same configuration as the structures tested in the present paper amounted to 0.99% and the difference with colourful panels was statistically irrelevant (df = 1, $\chi^2$ = 0.58, p = 0.45).

## 6.3. Limitations

The present study was based on young subjects familiar with the computer software, so it is unclear whether similar results would be obtained in the case of novice users. It should not be forgotten that the findings relate to fairly simple tasks involving both visual search and the computer mouse movement where the given layouts are visible solely during the selection process. The research was also restricted to the 'point and click' method and no other than square buttons with Latin characters and Arabic numbers were taken into account. One should also be cautious at drawing general conclusions about the role of preattentive visual information processing in the context of HCI task efficiency for the sake of limited number of colours used and only two types of panel configurations. The demonstrated experiments were carried out in the laboratory environment on the structures that are not used in computer programs and it is not known how the user will perform in a real environment.

## 6.4. Future research

The presented outcomes should be considered as a basis for further empirical investigations of the nature of the various attention deployment attributes in human computer communication. One may verify if and to what extent, other well known preattentive features have their importance in the panel operation efficiency. Subsequent research may also include other graphical objects, configurations, panel locations, comparison of beginners and advanced computer users. In light of the earlier research (e.g., Gramopadhye et al., 2002; Nickles III et al., 2003) it seems worthwhile to include the training factor in early vision search studies. Similar experiments can be as well conducted by means of other pointing devices or touch screens. Apart from the analysis of objective empirical data (acquisition times, errors) it is interesting to learn what are the subjective ratings of the individual panel layouts. And finally, the eye tracking systems and functional magnetic resonance imaging devices could substantially broaden the analysis.

## 7. Conclusions

The presented studies confirm that the preattentive visual processing mechanisms may play an important role and should be explicitly incorporated into the HCI field. One may argue that these results are trivial and could be easily foreseen since one could presume that the principal results obtained in the classical visual search research would apply also in the specific HCI environment. Nevertheless, we are convinced that such hypotheses should be formally verified given the specific conditions that included simultaneously panel visual search and visually controlled motor activities typical for direct manipulation. The obtained results partially confirm our scepticism, since it occurred that object background colour variable was significant only for horizontal panels. It means that the orientation of dimensionally restricted panels can considerably modify the colour visual preattentive processing. From a practical point of view, if designers want to take full advantage of the colour early vision process in graphical user interfaces they should rather avoid vertical layouts. We also proved that the colour preattentive visual processing is less salient in structures considerably spatially diversified, so such arrangements rather should not be applied in practice. Additionally, for the sake of higher search and click efficiency the practitioners should consider using higher colour contrasts.

The results obtained in this study seem to be giving a good reason for conducting more research combining standard human–computer activities with early vision processes. The further investigations in this area may lead to supplementing the existing graphical interface design guidelines, help in understanding the nature of HCI, and contribute to improvements in comprehensive cognitive models (e.g., Card et al., 1983; Newell, 1990; Anderson, 1993; Anderson et al, 1997; Kieras and Meyer, 1997; Byrne, 2001) that were proposed in the domain of human–computer interaction.